\documentclass[aps,pra,superscriptaddress,11pt]{revtex4} 
\usepackage{amsmath,amssymb,graphicx,times}
\begin{document}
\title{Quantum probes for ohmic environments at thermal equilibrium}
\author{Fahimeh Salari Sehdaran}
\affiliation{Faculty of Physics, Shahid 
Bahonar University of 
Kerman, Kerman, Iran}
\author{Matteo Bina, Claudia Benedetti, Matteo G. A. Paris}
\affiliation{Quantum Technology Lab, Dipartimento di Fisica {\em Aldo Pontremoli}, 
Universit\`a di Milano, I-20133 Milano, Italy}
\begin{abstract}
It is often the case that the environment of a quantum system
may be described as a bath of oscillators with Ohmic density
of states. In turn, the precise characterization of these
classes of environments is a crucial tool to engineer decoherence
or to tailor quantum information protocols. Recently, the use
of quantum probes in characterizing Ohmic environments at
zero-temperature has been discussed, showing that a
single qubit provides precise estimation of the cutoff frequency. On
the other hand, thermal noise often spoil quantum probing schemes,
and for this reason we here extend the analysis to complex
system at thermal equilibrium. In particular, we discuss the interplay
between thermal fluctuations and time evolution in determining the
precision {attainable by} quantum probes. Our results show that the presence
of thermal fluctuations degrades the precision for low values
of the cutoff frequency, i.e. values of the order $\omega_c 
\lesssim T$ (in natural units). For larger values of $\omega_c$
decoherence is mostly due to the structure of environment, rather
than thermal fluctuations, such that quantum probing by a single
qubit is still an effective estimation procedure.
\end{abstract}
\maketitle
\section{Introduction}
In the last decade, technological advances in control and 
manipulation of quantum systems have made quantum probes 
available to the characterization of a large set of physical 
platforms. In turn, a radically new approach emerged to probe 
complex quantum systems, which is based on the quantification 
and optimisation of the information that can be extracted by 
an immersed quantum probe as opposed to a classical one
\cite{breuer2002theory,Palma1996,Benedetti2018,Bina2018,Elliott,Streif,Troiani,Cosco,Benedetti2014}. 
Quantum probes offer two main advantages: on one hand, they 
often provide enhanced precision, due to the inherent sensitivity
of quantum system to environment-induced decoherence. On the other
hand, they provide non invasive techniques in order to
estimate parameters of interest, without perturbing too much
the system under investigation.
\par
In this paper, we address the use of the simplest quantum probe, 
a single qubit, {as it already embodies all the 
desired  properties of an effective probe: it is small, only weakly 
invasive, and it can be easily manipulated and controlled 
\cite{Zhang08, Berkley13, Lolli15}.}
Our aim is to characterize the spectral properties 
of a bath of oscillators, which itself provides a quite general 
model, suitable to describe several complex systems of interest 
for quantum information science and reservoir engineering \cite{Paavola,Martinazzo,Myatt,PiiloManiscalco,TamascelliSmirne,TamascelliLemmer}. In particular, we focus on the 
cutoff frequency {$\omega_c$} of the environment, which is linked to the 
environment correlation time and, in turn, to the available coherence 
time for communication and computation. Indeed, a precise characterization 
of the spectral density is a crucial step to the engineering of reservoirs,
tailored to specific tasks. Recently, the effective use of 
a single qubit quantum probe to characterize Ohmic environments 
at zero-temperature has been analyzed and discussed \cite{Benedetti2018}.
On the other hand, thermal fluctuations often spoil the effectiveness 
of quantum metrological protocols, the most dramatic case being represented
by quantum interferometry, where an infinitesimal amount of noise is enough
to kill Heisenberg scaling and reinstates the shot noise limit \cite{rqi}.
{{In turn, the effect of temperature has been 
analyzed in different metrological contexts, for example the out-of-equilibrium 
regimes \cite{cavina18} and phase estimation in Gaussian states \cite{garbe19}.}}
For these reasons, we extend here the analysis to the more realistic case 
of complex systems at thermal equilibrium, and discuss in details the 
interplay among thermal fluctuations and time evolution in making the 
qubit an effective probe for the cutoff frequency of its environment.  
In this context, a relevant feature of our probing technique is the 
pure dephasing
nature of the interaction between the qubit and its environment.
This means that while the Ohmic system has a fixed temperature, 
the probe has access to the full set of out-of-equilibrium states \cite{qcse}, while not exchanging energy with the Ohmic system. 
As we will 
see, this provides room to optimise the probing strategy and to enhance 
sensitivity over classical (thermal) probes. 
\par
Any probing strategy requires control of the initial state of the probing 
system, as well as of the coupling with the probed one. Concerning the 
detection of the probe after interaction, we exploit results from local 
quantum estimation theory (QET), which provides the necessary tools to 
determine the most informative measurement and the most precise estimator
and, in turn, to optimise the extraction of information from the 
quantum probe \cite{Paris2009}. Indeed, QET has been effectively employed 
in different contexts \cite{Benedetti_PLA,Zwick_Kurizki,Monras_EstLoss,Fujiwara_QChannel,Fujiwara_EstPauli,Pinel_Gaussian,Brida_EstEnt_PRL,Brida_EstEnt,Blandino_HomoDiscord,Benedetti_Discord,Genoni_PhaseDiff,Monras,Bina_PhasePNR,Walmsley_ExpEst,Spagnolo_PhaseEst,Brunelli_Thermo,Correa_Thermo,Stenberg14,Bina_Dicke,Rossi_Diamagnetic,Tamascelli_Feynman}, in order to individuate the most convenient detection scheme, and 
to evaluate the ultimate quantum bounds to precision.
In this work, we address the characterization of Ohmic environments 
at thermal equilibrium, i.e. the estimation of their cutoff frequency, assuming 
that the nature of the environment is known, i.e. the value of the ohmicity 
parameter. On the other hand, we optimize the strategy over the initial 
preparation of the probe qubit,
the interaction time and the detection scheme at the output. In particular,
we pay attention to the overall estimability of the cutoff frequency, as 
measured by the quantum signal-to-noise ratio, in different temperature
regimes. As we will see, the presence of thermal fluctuations degrades 
the estimation precision. On the other hand, the negative effects of 
temperature are relevant only for small values of the cutoff frequency, 
i.e. values of the order $\omega_c \lesssim T$ (in natural units). For larger
values of the cutoff frequency the decoherence of the probe is mostly due
to the structure of the environment, rather than thermal fluctuations, 
and therefore the overall estimation procedure is still very effective, 
with performances very close to the zero temperature case.
\par 
The paper is structured as follows. In Section \ref{s:model}, 
we describe the interaction model, establish notation, and briefly 
review the ideas and the tools of QET. In Sec. 
III we present our results and discuss in details the interplay 
between thermal fluctuations  and time evolution in determining the precision 
of quantum probes. Section IV closes the paper with some concluding 
remarks.
\section{The model}\label{s:model}
Our quantum probe is a single qubit with energy gap $\omega_0$, 
which interacts with a bosonic reservoir at thermal equilibrium. 
The total  Hamiltonian may be written as
\begin{equation}\label{htot}
{\cal H}=\frac{\omega_0}{2} \sigma_{3} +\sum_{k}\omega_{k}\,b^{\dagger}_{k}
\,b_{k}+\sigma_{3}\, \sum_{k}(g_{k}\,b^{\dagger}_{k}+g^{*}_{k}\,b_{k})\,,
\end{equation}
where $\omega_{k}$ is the  frequency of the $k$-th reservoir mode 
and we use natural units with $\hbar=k_{B}=1$. The Pauli matrix $\sigma_{3}$ acts
on the qubit and $[b_{k},b^{\dagger}_{k}]=\delta_{k\,k^{'}}$ describes the modes
of the bath. The $g_{k}$'s are coupling constants, describing the interaction 
of each mode with the qubit. Their distribution is usually described in terms
of the so-called spectral density of the bath, which is defined as 
$J(\omega)=\sum_{k}\,|g_{k}|^{2}\,\delta(\omega_{k}-\omega)$. The spectral density
depends on the the specific features of the physical system and may be often engineered in order to enable control of quantum  decoherence \cite{Palma1996}.
The model described by ${\cal H}$ in Eq. (\ref{htot})
is exactly solvable, making it possible to analyze the mechanisms creating 
entanglement between the qubit and environment, which in turn is at the core 
of decoherence processes \cite{breuer2002theory,Palma1996}. 
\par 
We are interested in probing properties of the environment by performing measurements
on the qubit. To this aim, we study the reduced dynamics of the qubit assuming that
the environment is at thermal equilibrium, i.e. $$\rho_{\hbox{\tiny{E}}}
=\frac1Z \exp\left\{-\frac1T \sum_{k}\omega_{k}\,b^{\dagger}_{k} \,b_{k}\right\}\,,$$ 
where $Z=\hbox{Tr}\left[\exp\{-\frac1T \sum_{k}\omega_{k}\,b^{\dagger}_{k} \,b_{k}\}\right]$ is the partition function 
and $T$ denotes the temperature.
In particular, our goal is to probe the cutoff frequency of Ohmic environments, 
i.e. the quantity $\omega_c$ appearing in spectral densities of the form 
\begin{equation}
J_s(\omega,\omega_c)= \omega_c \left(\frac{\omega}{\omega_c}\right)^s 
\exp\left\{-\frac{\omega}{\omega_c}\right\}\,.
\label{homspec}
\end{equation}
The cutoff frequency is a crucial parameter for applications
in quantum information science, since it is linked to the 
environment correlation time and, in turn, to the available coherence 
time for communication and computation. The quantity $s$ is a real 
positive number, which instead governs the behaviour of
the spectral density at low frequencies. Upon varying $s$ we move 
from the sub-Ohmic regime ($s<1$), to Ohmic ($s=1$), and to super-Ohmic 
one ($s>1$). In the following, in order to make some explicit quantitative 
statements, we will often refer to the paradigmatic values $s=0.5, 1, 3$ {\cite{Leggett,Shnirman}}.
\par
\begin{figure}[h!]
\includegraphics[width=0.48\columnwidth]{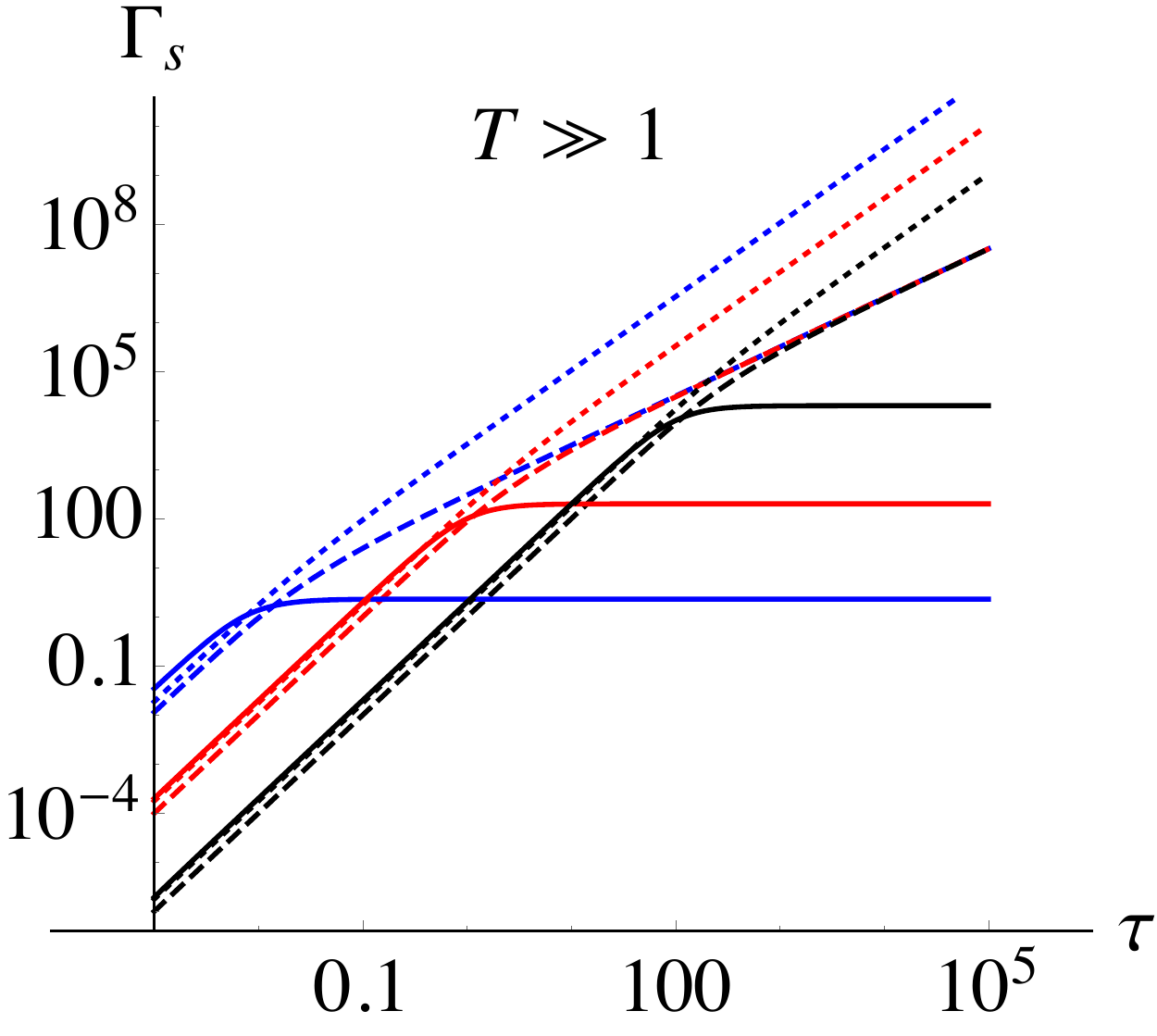}
\includegraphics[width=0.48\columnwidth]{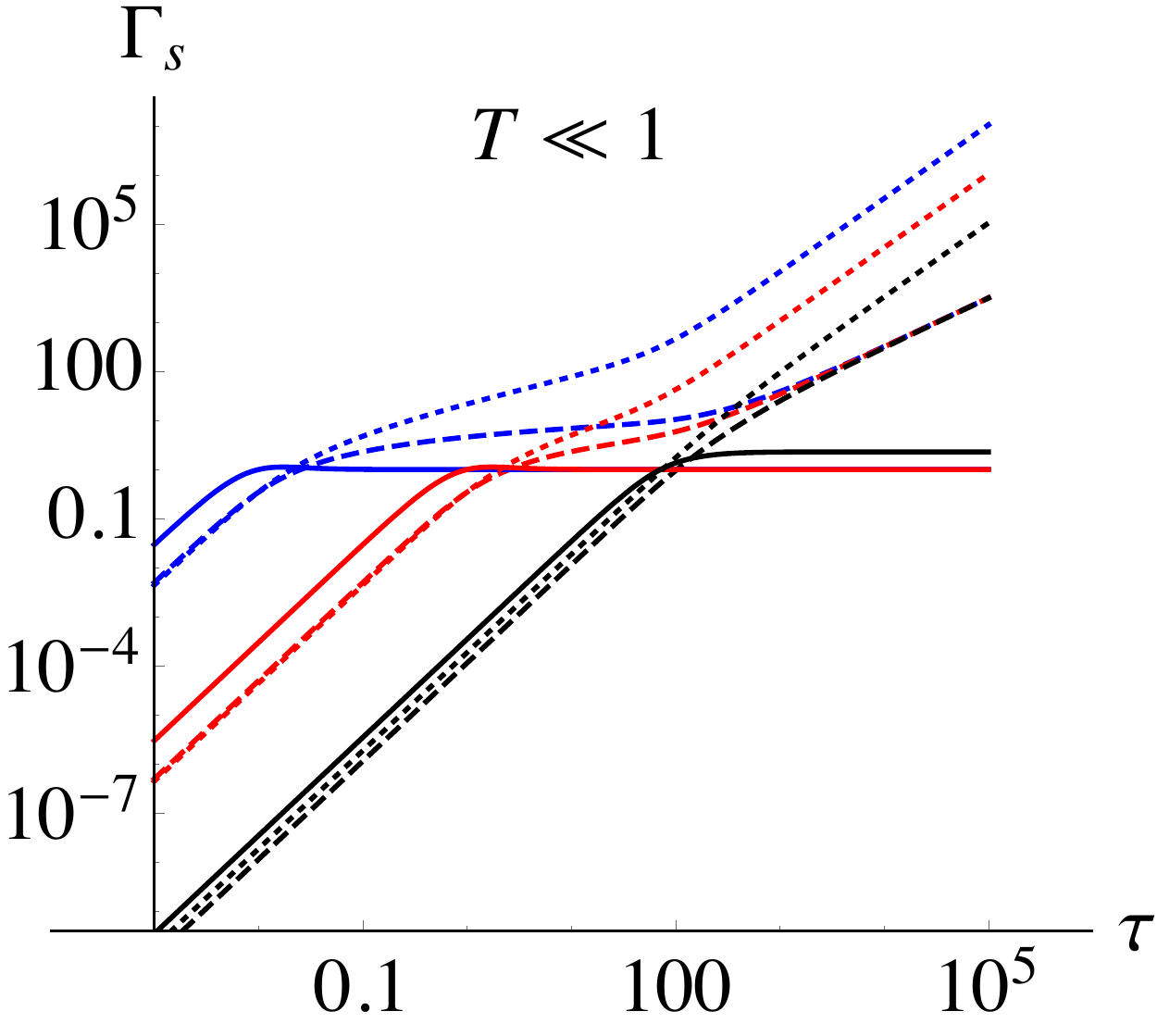}
\caption{\label{f:df}Decoherence function $\Gamma_s$ as a 
function of the dimensionless time $\tau$  for different temperatures, cutoff frequencies, and ohmicity
parameters. The left panel reports $\Gamma_s$ in the high 
temperature regime (the plot is for $T =10^2$), whereas the right panel 
shows it for low temperature, $T=10^{-2}$. In both plots, black lines are 
for $\omega_c=10^{-2}$, red ones for $\omega_c=1$ and blue ones for $\omega_c=10^2$.
Finally, solid lines denote results obtained for super-Ohmic environments ($s=3$), dashed for Ohmic ($s=1$) and dotted ones for sub-Ohmic ($s=0.5$).}
\end{figure}

The initial state of the combined system, qubit and environment, is described 
by the density matrix
\begin{equation}
\rho_{\hbox{\tiny{QE}}}(0)=\rho_{\hbox{\tiny{Q}}}(0)\otimes\rho_{\hbox{\tiny{E}}}
\end{equation}
where $\rho_{\hbox{\tiny{E}}}$ is given above. The initial preparation of the qubit probe 
  $\rho_{\hbox{\tiny{Q}}}(0)$ should be optimised in order to extract the maximum 
possible information on $\omega_c$ from measurements performed on the 
qubit after the interaction with the environment. This optimization has 
been performed in Ref.~\cite{Benedetti2018}  for environments at zero temperature. 
The proof does not depend on the structure of the environment, 
but only on 
the  pure dephasing map of the qubit. Since
the same dynamical map is considered here, the proof
 holds also for thermal environments, and thus we consider 
$\rho_{\hbox{\tiny{Q}}}(0)=|+\rangle\langle +| = \frac12\, ({\mathbb I} + \sigma_1)$, 
where $|+\rangle=\frac{1}{\sqrt{2}}
\,(|0\rangle+|1\rangle)$, being {$\{|0\rangle,|1\rangle\}$} the
computational basis, i.e. the eigenstates of $\sigma_3$.
We now move to the interaction picture, where the Hamiltonian and the evolution operator take on the expressions:
\begin{align}
{\cal H}_I &= \sigma_3\sum_k\left(g_k b_k^{\dagger}e^{i\omega_k \tau}+g_k^{*}b_ke^{-i\omega_k \tau}\right)
\\
U_I(\tau)&\propto \exp\left[\frac12 \sigma_3\sum_k\left(\alpha_k b_k^{\dagger}-\alpha_k^{*}b_k\right)\right]
\end{align}
where $\alpha_k=2g_k\frac{1-e^{i\omega_k \tau}}{\omega_k}$ \cite{breuer2002theory}. 
If we assume a continuum of environment's modes, we can use the spectral density \eqref{homspec} to
evaluate  the evolved state of the qubit probe upon tracing out the
environment {$\rho_{\hbox{\tiny{Q}}}(\tau)=\hbox{Tr}_{\hbox{\tiny{E}}}\left[U_I(\tau)\,\rho_{\hbox{\tiny{QE}}}(0)
\,U_I^\dagger(\tau)\right]$, which explicitly reads:}
\begin{align}
\label{rqt}
\rho_{\hbox{\tiny{Q}}} (\tau)  & = 
\frac12 \left( {\mathbb I} + e^{-\Gamma_s(\tau,T,\omega_c) } \sigma_1\right) 
\,,
\end{align}
where
\begin{align} \label{gamman}
\Gamma_{s}(\tau,T,\omega_c) 
= \!\!\! \int_0^{\infty}\!\!\!\!d\omega \, J_s(\omega,\omega_c)\, 
\frac{1-\cos \omega \tau}{\omega^2}\coth\left(\frac{\omega}{2T}\right)\,,
\end{align}
is usually referred to as the {\it decoherence function}, with $\exp\{-\Gamma_s (\tau,T,\omega_c) \}$ 
referred to as the decoherence factor. 
{Notice that in Eq. (\ref{gamman}) frequencies, time and temperature 
are dimensionless quantities expressed in units of the probe frequency $\omega_0$}. The decoherence function depends on the temperature $T$
of the environment, and on the form of the spectral density 
{$J_s(\omega,\omega_c)$ \cite{breuer2002theory,Benedetti2018}, 
i.e. on the cutoff frequency $\omega_c$}  and the ohmicity parameter $s$. 
{An analytic form of the integral in Eq.~(\ref{gamman}) may be 
obtained at any temperature, time and Ohmicity parameter by 
expanding the hyperbolic cotangent    $\coth(x)=1+2\sum_{n=1}^\infty {\rm e}^{-2 n x}$. Then, the decoherence function may be written as 
$$\Gamma_{s}(\tau,T,\omega_c) = \Gamma_{s}(\tau,0,\omega_c) +2\sum_{n=1}^\infty\left(\frac{T}{T+n\,\omega_c}\right)^{s-1}\Gamma_{s}\left(\tau,0,\frac{T \omega_c}{T+n\,\omega_c}\right)\,, $$ 
which explicitly reads
\begin{equation}\label{GammaGen}
\Gamma_{s}(\tau,T,\omega_c) \!=\!\Gamma_{s}(\tau,0,\omega_c)+ s(s-1)\left (\frac{T}{\omega_c}\right)^{s-1}\!\!\frac{\Gamma_e[s-1]^2}{\Gamma_e[s+1]} F(\zeta)
\end{equation}
where $\Gamma_e[z]=\int_0^\infty\!dt\, t^{z-1} e^{-t}$ is the Euler 
Gamma function and where we introduced the function
\begin{equation}
F(\zeta )\!\equiv \!2\zeta[s-\!1,1\!+\!\text{Re}(w)]-\zeta[s-\!1,1\!+\!w] - \zeta[s-\!1,1\!+\!w^*] \, ,
\end{equation}
where $w\equiv T \,\omega_c^{-1}+i \, T\,\tau$, 
$\zeta[p,q]=\sum_{k=0}^\infty(k+q)^{-p}$ is the 
generalised (Hurwitz) Zeta function and $\Gamma_{s}(\tau,0,\omega_c)$ is the 
decoherence function at zero temperature, i.e. \cite{Benedetti2018} 
\begin{equation}
\Gamma_{s}(\tau,0,\omega_c)\!=\! \Gamma_e[s\!-\!1]\!\left \{\! 1\!-\!\frac{\cos\big[ (s\!-\!1)\arctan(\omega_c \tau) \big]}{\left ( 1+\omega_c^2 \,\tau^2 \right )^{\frac{s-1}{2}}}\!\!\right \} \!.
\end{equation}}
The behaviour of the decoherence function, {that from now on we denote as $\Gamma_s\equiv \Gamma_{s}(\tau,T,\omega_c)$, as a function of 
the dimensionless time $\tau$ is shown in Fig. \ref{f:df}, for different cutoff frequencies, Ohmicity parameters and two regimes of high and low 
temperature of the environment}.
As it is apparent from the plots, for short times {($\tau \ll 1$)} the decoherence function 
{follows a power-law scaling} for any value of the other parameters. 
More precisely, {from a first-order approximation, it scales as $\tau^2$:
\begin{equation}
\Gamma_s \, {\simeq} \, \frac12\,\omega_c^{2} \Gamma_e(s-1)\!  \left[2\left(\frac{T}{\omega_c} \right )^{s+1}\!\!\!\!\!\zeta\left(\!s\!+\!1,\frac{T}{\omega_c}\right)-1  \right] \tau^2 \, .
\end{equation}}
{The asymptotic behaviour at long times, instead, is different for the three choices of the Ohmicity parameters. In particular, in the super-Ohmic case with $s=3$ the decoherence function saturates to a constant value, at any temperature $T$. This means that the stationary state of the qubit is not a fully dephased one, and that the residual degree of coherence is 
larger for values of the parameters leading to smaller saturation values 
of $\Gamma_3$. In the other cases, sub-Ohmic with $s=0.5$ and Ohmic with $s=1$, the decoherence function scales, respectively, as $\Gamma_{0.5} \sim \tau^\frac32$ and $\Gamma_1\sim \tau$, meaning that the stationary state of the qubit probe has been completely decohered. The long-time behavior of the decoherence function is also important from the point of view of the characterization of the type of Ohmic-like environment, namely the asymptotic scaling clearly distinguishes and characterizes the Ohmicity parameter of the considered structured reservoir.}
\subsection{Quantum parameter estimation}
\label{s:lqe}
The density matrix $\rho_{\hbox{\tiny{Q}}}(\tau,\omega_c,s,T)$
in Eq. (\ref{rqt}) describes the state of the qubit probe
after the interaction with the environment. As such, it depends on the 
interaction time $\tau$, which is basically a free parameter, on the 
temperature $T$ and the Ohmicity parameter $s$, which are fixed 
by the experimental conditions, and on the cutoff frequency $\omega_c$ 
of the environment, which is the parameter we would like to estimate.
In the jargon of quantum estimation, it is usually referred to as a
{\it quantum statistical model}.
According to this classification, and in order to simplify the notation, 
in this Section we will use the following shorthands
\begin{align}
\rho_{\hbox{\tiny{Q}}}(\tau,\omega_c,s,T) & \longrightarrow \rho_c \qquad\qquad
\frac{\partial}{\partial \omega_c}  \longrightarrow \partial_c\notag\,.
\end{align}
Our task is to optimize the inference of $\omega_c$ by performing measurements
on $\rho_c$. To this aim, we employ results from quantum estimation theory \cite{Paris2009}, which provides tools to find the best detection scheme and
to evaluate the corresponding lower bounds to precision. We assume that the 
value of the temperature $T$ and the Ohmicity parameter $s$ are fixed,
whereas the value of interaction time is a free parameter, over which we may
further optimize the precision.
\par
Let us consider the family of quantum states $\rho_c$, which is labeled by 
 the cutoff frequency $\omega_{c}$. In order to estimate $\omega_c$ we perform 
measurements on repeated preparations of the quantum probe and then process 
the overall sample of outcomes. 
Let us denote by $X$ the observable 
measured on the probe, and by $p(x|c)$ the conditional distribution of its 
outcomes when the true value of the cutoff frequency is $\omega_c$. We also
denote by $M$ the number of repeated measurements. Once 
$X$ is chosen and a set of outcomes ${\mathbf x}=\{x_1,...,x_M\}$ is collected,
we process the data by an {\it estimator} $\hat \omega_c \equiv \hat 
\omega_c ({\mathbf x})$, i.e. a function from the space of datasets to the manifold
of the parameter values. The {\it estimate} of the cutoff frequency is the average
value of the estimator over data, whereas the {\it precision} of this estimate corresponds to the variance of the estimator i.e. 
\begin{align}
\overline{\omega}_c = \int\!\! d{\mathbf x}\, p(\mathbf{x}|c) 
\, \hat \omega_c ({\mathbf x})\,, \qquad
V_c \equiv \hbox{Var}\, \omega_c = 
\int\!\! d{\mathbf x}\, p(\mathbf{x}|c) 
\, \Big[\hat\omega_c ({\mathbf x})- \overline{\omega}_c  \Big]^2\,,
\end{align}
where $ p(\mathbf{x}|c) = \Pi_{k=1}^M\, p(x_k|c)$, 
since the repeated measurements are independent on each other. 
The smaller is $V_c$, the more precise is the estimator. In fact, 
there is a bound to the precision of any unbiased estimator (those satisfying
the condition $\overline{\omega}_c \rightarrow \omega_c$ for $M \gg1$), given by 
the Cram\'er-Rao (CR) inequality:
\begin{equation}
V_c \ge \frac{1}{M F_c}\,, \qquad F_c = \int\!\! dx\, p(x|c) 
\Big[\partial_c \log p(x|c)\Big]^2\,,
\end{equation}
where $F_c$ is the (single-measurement) Fisher information (FI).
The best, i.e. more precise, measurement to infer the value of 
$\omega_c$ is the measurement maximising the FI, where the maximization
should be performed over all the possible observables of the probe. 
To this aim, one introduces the symmetric logarithmic derivative 
$L_{\omega_c}\equiv L_c$ (SLD), as the operator which
satisfies the relation
\begin{equation}
L_c\,\rho_{c}+\rho_{c}\,L_{c}=2\partial_{c} \rho_{c}\,.
\end{equation}
The quantum CR theorem states that the optimal quantum measurements are 
those corresponding to the spectral measure of the SLD, and consequently 
$F_c \leq H_c = \hbox{Tr}[\rho_c\,L_c^2]$, where $H_c$ is usually 
referred to as the quantum Fisher information (QFI). The quantum CR 
inequality then follows
\begin{equation}\label{qcr}
V_c \ge \frac{1}{M H_c}\,
\end{equation}
and it represents the ultimate bound to precision, taking into account both 
the intrinsic (quantum), and extrinsic (statistical), source of fluctuations 
for the estimator.
Starting from the diagonal form of the quantum statistical model 
$\rho_c=\sum_n \rho_{n}\,\big|\phi_{n}\rangle\langle\phi_{n}\big|$, 
where both the eigenvalues and the eigenvectors do, in general, depend
on the parameter of interest, we arrive at a convenient form of the 
QFI
\begin{equation}\label{qfidm}
H_c=\sum_n\frac{(\partial_c \rho_n)^2}{\rho_n}+ 
2\sum_{n\ne m}\frac{(\rho_n-\rho_m)^2}{\rho_n+\rho_m}\,\big|\langle\phi_m | 
\partial_c\, \phi_{n}\rangle\big|^2\,
\end{equation}
where, for our qubit case, $n,m=1,2$.
The first term in Eq.~(\ref{qfidm}) is the FI of the distribution
of the eigenvalues $\rho_{n}$, whereas the second term is a positive
definite, genuinely quantum, contribution, explicitly quantifying 
the potential quantum enhancement of precision. 
{Any measurement $X$ on the system is associated to its FI, and different measurements lead to different degrees of precisions through the CR bound. Hoever, when a measurement is found, such that }
 the condition $F_c=H_c$ is satisfied, the measurement is said to 
be {\it optimal}. If the equality in Eq.~(\ref{qcr}) is satisfied the
corresponding estimator is said to be {\it efficient}.
A global measure of the estimability of a parameter, weighting the variance
with the value of the parameter, is given by the signal-to-noise ratio 
$R_c=\omega_c^2/V_c$. The quantum CR bound may then be rewritten in terms 
of $R_c$ as follows
\begin{align}\label{qsnr}
R_c \leq Q_c = \omega_c^2 H_c\,,
\end{align}
where $Q_c$ is referred to as the quantum signal-to-noise ratio (QSNR),
and itself represents the ultimate quantum
bound to the estimability of a parameter \cite{Benedetti2014,Paris2009}.
The larger is the QSNR, the (potentially) more effective is the estimation 
scheme \cite{Benedetti2018}.
Here "potentially" refers to the fact that a large value of the 
QSNR means a large QFI, which in turn tells us about
the maximum precision that can be achieved. However, it does not 
say anything about the best estimator that must be employed in order to 
process the output data and to infer the value of the parameter. A large $Q_c$ is a necessary step in order to precisely estimate the parameter. 
Finally, we notice that $\omega_c$ takes value on a subset of the real axis and this means that even if the optimal measurement does depend on the value to be estimated, the ultimate precision dictated by the 
quantum Cramer-Rao bound may be achieved by a two-stage adaptive 
scheme \cite{bng00}.
\section{\label{sec:level1} Quantum probes for Ohmic environments at 
thermal equlibrium}
In this section, using results of Section \ref{s:model}, 
we discuss the performances of a qubit probe in 
estimating the cutoff frequency of Ohmic environments at 
thermal equlibrium. Our starting point is 
the state of the probe after the interaction with the environment,
which provides the quantum statistical model 
$\rho_c$. 
We assume that the temperature $T$ and the Ohmicity parameter 
$s$ are fixed by the experimental conditions, whereas the interaction 
time $\tau$ may be tuned in order to maximise the quantum Fisher information 
$H_c$ and, in turn, the quantum signal-to-noise ratio $Q_c$.
To this aim, we first diagonalize $\rho_c$ and then use Eq. (\ref{qfidm}). After some
algebra, we arrive at 
\begin{align}\label{hctau}
H_c(\tau)=\frac{\big[\partial_c
\Gamma_s\big]^2}{\exp\left[2\Gamma_s\right]-1}\,,
\end{align}
where we have omitted the explicit dependence on $T$ {and $\Gamma_s$ is given by the explicit analytic formula (\ref{GammaGen})}. Starting
from Eq. (\ref{hctau}) we have maximised 
$Q_c(\tau) = \omega_c^2\, H_c(\tau)$ over 
the interaction time $\tau$ at different fixed values of $T$ 
and $s$. In particular, we have considered three specific values 
of $s=0.5,1,3$ in order to address sub-Ohmic, Ohmic, and 
super-Ohmic regimes. 
\par
\begin{figure*}[h!]
\includegraphics[width=0.32\textwidth]{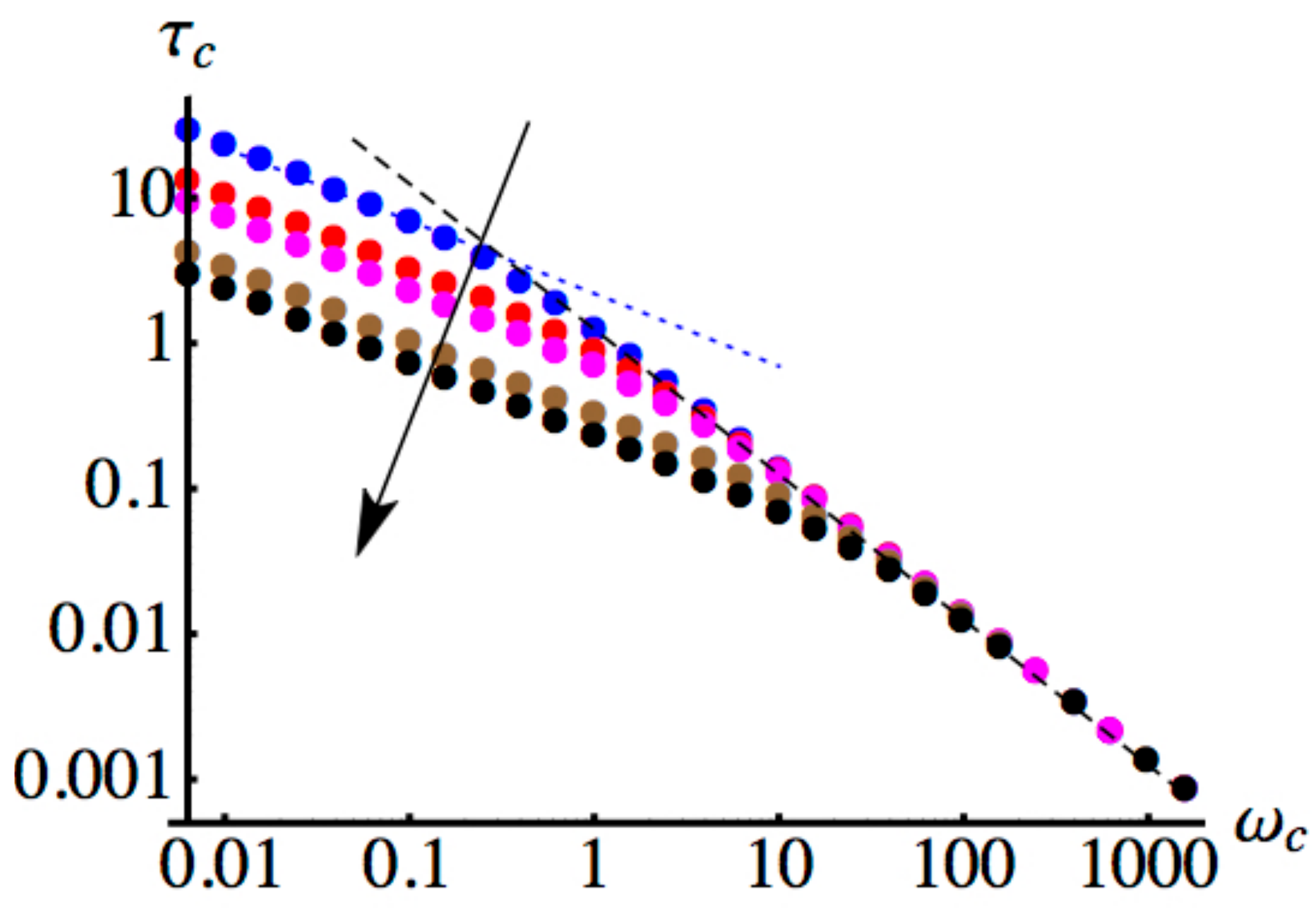}
\includegraphics[width=0.32\textwidth]{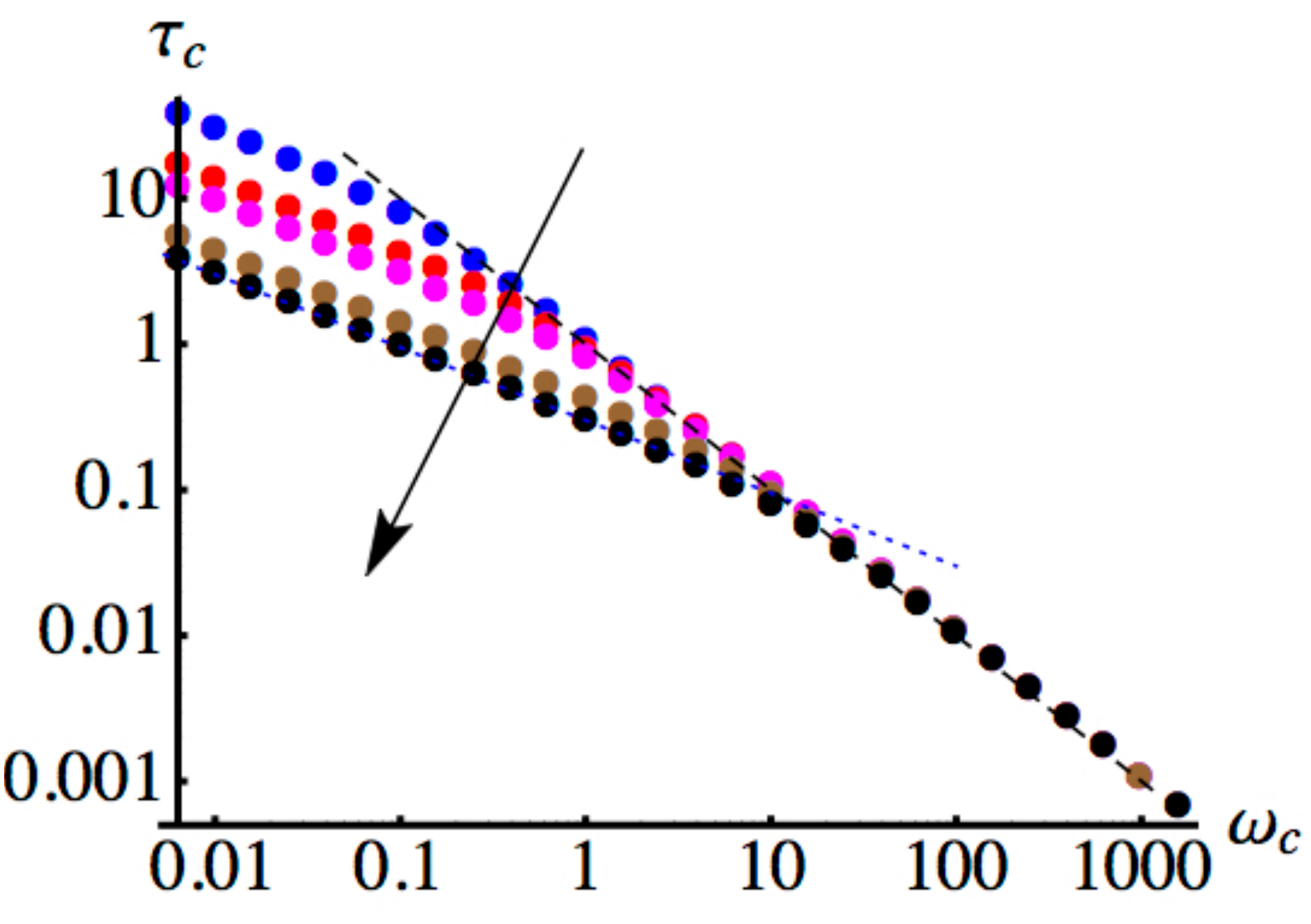}
\includegraphics[width=0.32\textwidth]{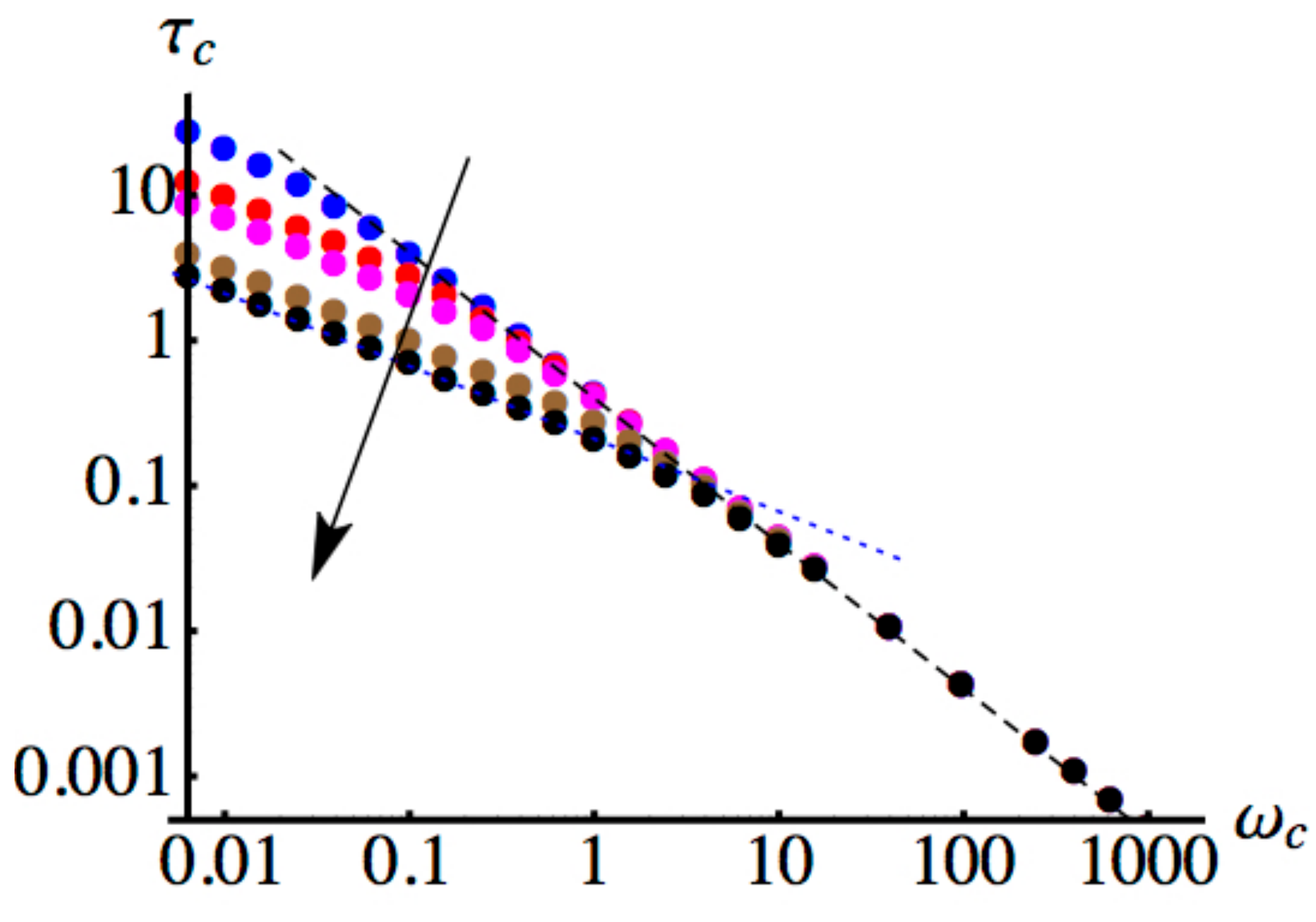} \\
\includegraphics[width=0.32\textwidth]{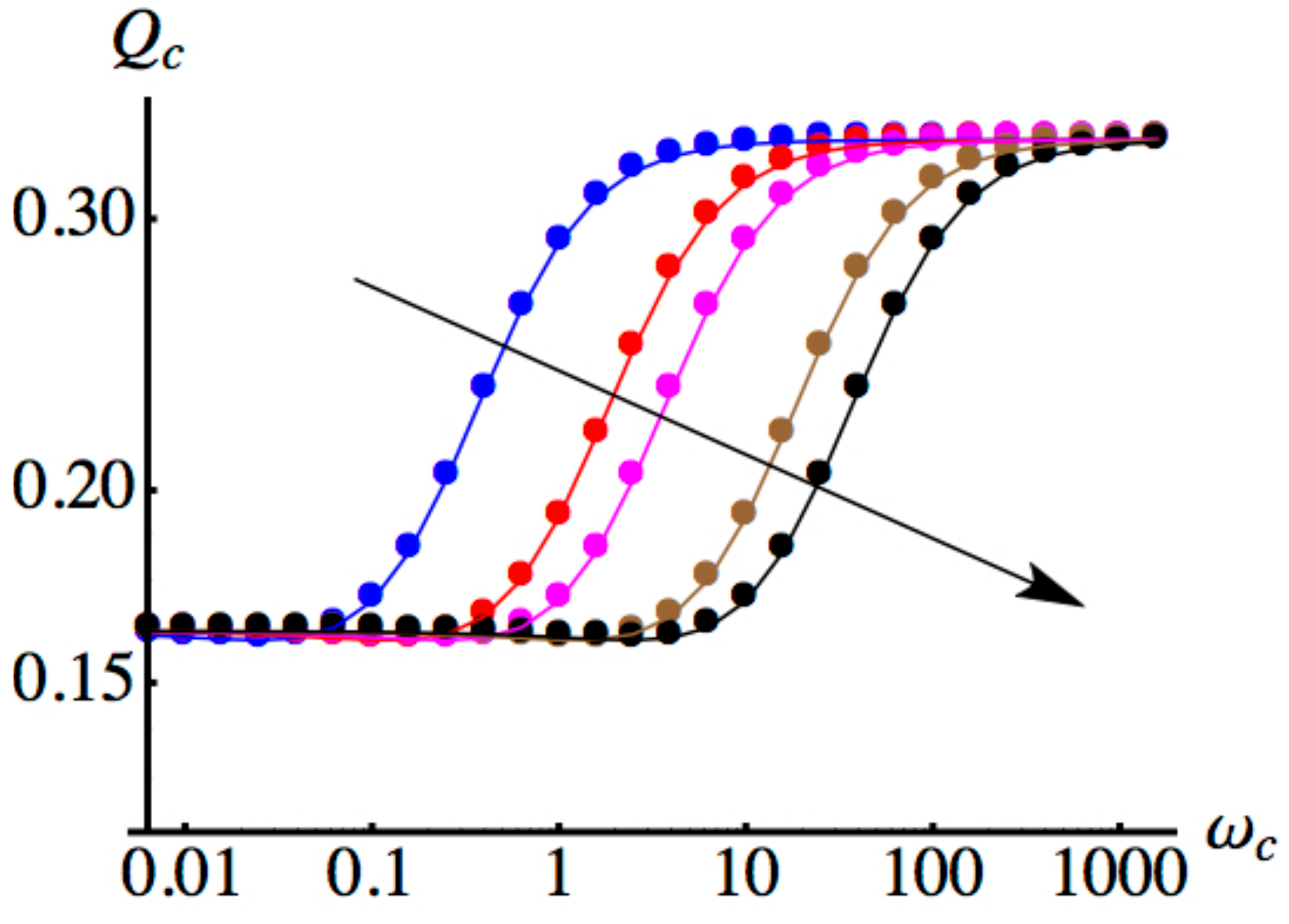}
\includegraphics[width=0.32\textwidth]{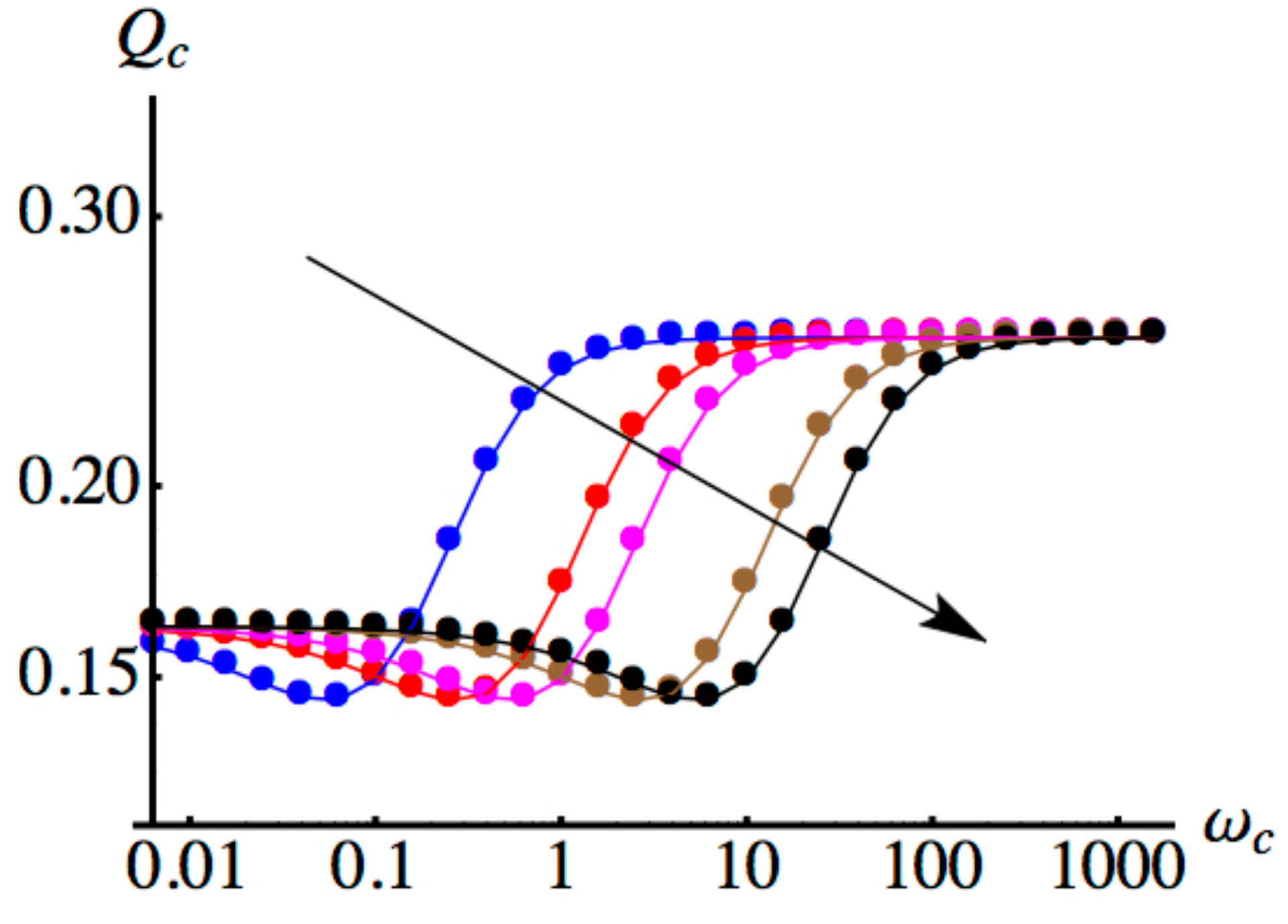}
\includegraphics[width=0.32\textwidth]{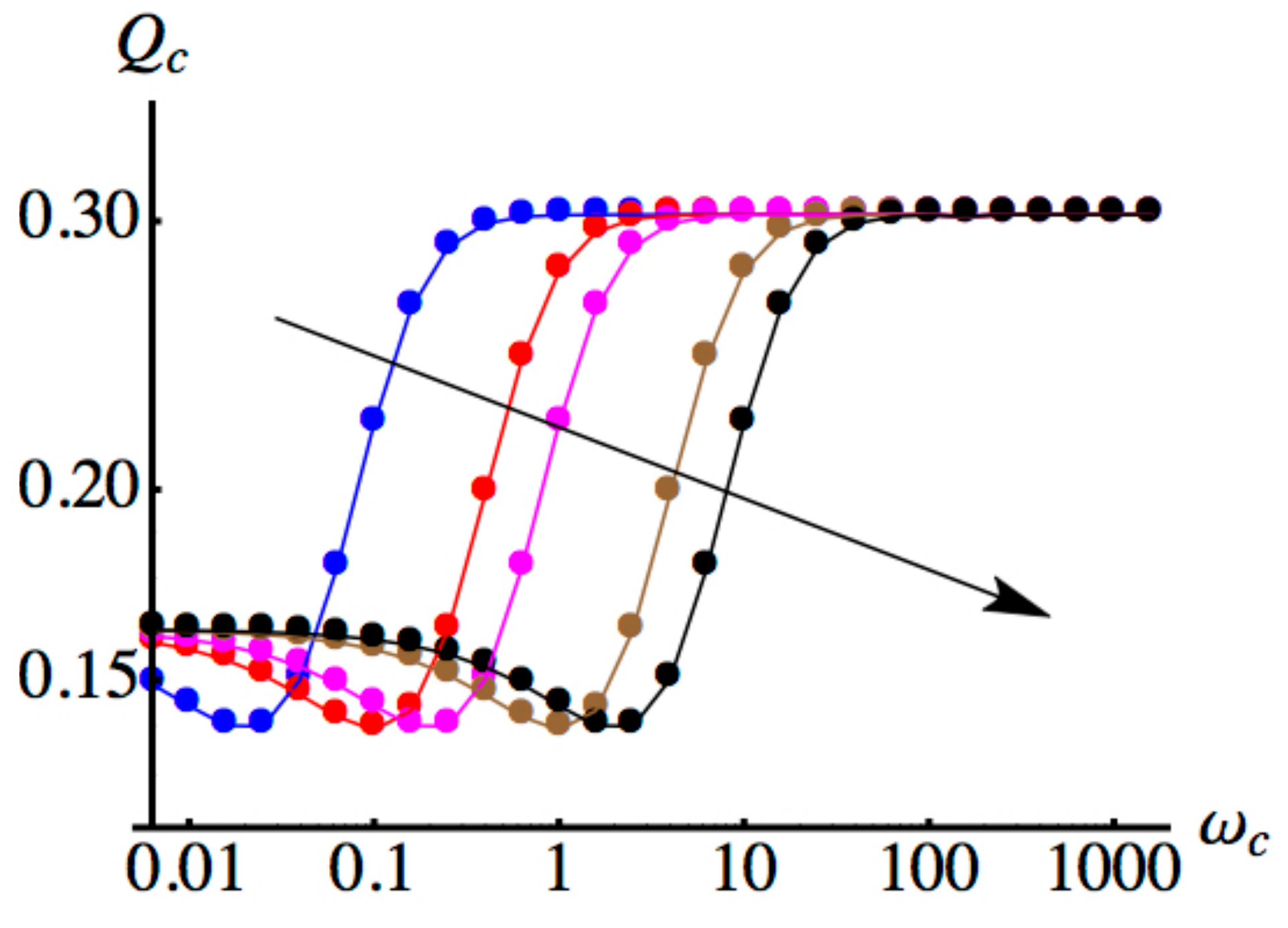}
\caption{Upper plots: the optimal interaction time $\tau_c$ as
a function of the cutoff frequency $\omega_c$ for different values of the
temperature (from top to bottom we have $T=0.1, 0.5, 1.0, 
5.0, 10.0$, arrows point to increasing temperature). 
From left to right the plots refer to $s=0.5, 1, 3$. {Dashed lines indicate the scaling of $\tau_c$ with $\omega_c$ in the two regimes of low and high cutoff frequency.} 
Lower plots: the optimised values of the QSNR $Q_c$, achieved
for the interaction times of the upper plots, as a function
of the cutoff frequency for different values of the
temperature (from top to bottom we have 
$T=0.1, 0.5, 1.0, 5.0, 10.0$, arrows point to 
increasing temperature). From left to right the 
plots refer to $s=0.5, 1, 3$. }
\label{f2res}
\end{figure*}

In Fig. \ref{f2res} we show the results of the optimisation.
The upper plots show the optimal interaction time $\tau_c$ as
a function of the cutoff frequency for the three considered 
values of the Ohmicity parameter, and for different values 
of temperature ($T= 0.1, 0.5, 1.0, 5.0, 10.0$), 
whereas the plots in the lower panels show the corresponding 
optimised values of the QSNR $Q_c$, for the same values 
of $s$ and $T$. In all the plots, the arrow denotes
increasing values of temperature. 
In the region of low
cutoff frequencies, the decoherence of the probe qubit is 
governed by thermal fluctuation, rather than the structure
of the environment. As a consequence, a larger interaction
time, scaling as $\tau_c \propto \omega_c^{-1/2}$, is needed
to imprint the maximal possible information about $\omega_c$ 
on the probe. The corresponding values of $Q_c$ are anyway 
smaller than those achievable in the zero temperature case{, which corresponds to the upper saturation level for $\omega_c \gg T$}. 
Upon increasing the cutoff frequency, the zero temperature 
scaling of the optimal time, $\tau_c \propto \omega_c^{-1}$ 
is recovered, as well as the values of the optimised QSNR.
Combining numerical results with Eq. (\ref{GammaGen}) we see
that for $\omega_c > T$ the optimal time scales as follows: 
$\tau_c \simeq \frac54\, \omega_c^{-1}$ for $s=0.5$, 
$\tau_c \simeq \omega_c^{-1}$ for $s=1$, 
and $\tau_c \simeq \frac25\, \omega_c^{-1}$ for $s=3$,
independently on the temperature itself.
\par
The transition from the regime of decoherence induced
by temperature to the regime of decoherence governed by
the structure of the environment may be traced back to 
the behaviour of the decoherence function $\Gamma_s$, 
and takes place for cutoff frequencies of the order 
$\omega_c \simeq T$. Remarkably, as far as $\omega_c$ 
is exceeding this threshold value, the value of the 
QSNR $Q_c$ quickly increases, and reach the zero 
temperature level independently on the temperature 
of the environment. We notice that even in the region
of low cutoff frequencies where thermal fluctuations 
degrade performances (the QSNR is reduced by a factor about 2/3), 
qubit probes are still providing information about
their environment.
\section{Conclusions}
In this paper, we have addressed estimation of the cutoff frequency 
of a complex Ohmic environment at thermal equilibrium. Our approach
is based on the use of a quantum probe, i.e. a simple quantum system
which undergoes decoherence due to its interaction with the environment.
In particular, we have focussed on the use of a single qubit subject 
to environment-induced dephasing, and have evaluated
the optimal interaction time between the probe and the environment, 
needed to imprint on the qubit the maximum information about 
the cutoff frequency. In addition, we have discussed the interplay
between thermal fluctuations and time evolution in determining the 
precision of quantum probes. 
\par
Our results shows that the presence of thermal fluctuations 
degrades the precision for low values of the cutoff frequency,
whereas for larger values a single qubit is still providing nearly 
optimal performances, i.e. precision close to the zero temperature
case. This behaviour may be explained in terms of the mechanisms 
responsible for the decoherence of the qubit. In the region of 
low cutoff frequencies, the decoherence of the probe is 
governed by thermal fluctuations, rather than the structure
of the environment. As a consequence, a larger interaction
time, scaling as $\tau_c \propto \omega_c^{-1/2}$, is needed
to imprint the maximal possible information about $\omega_c$ 
on the probe, and the corresponding values of the QSNR are  
smaller than those achievable in the zero temperature case. 
On the other hand, upon increasing the cutoff frequency, 
thermal fluctuations are no longer the main cause of decoherence, 
and the zero temperature scaling of the optimal interaction time, 
$\tau_c \propto \omega_c^{-1}$ is recovered, as well as the 
values of the optimised QSNR.  Our results pave the way for 
possible applications to realistic room temperature systems, 
as well as for estimation of more than a single parameter 
in system-environment couplings with general spectra. 
\acknowledgments{
The authors thank Luigi Seveso and Sholeh Razavian for several 
useful discussions.This work has been supported by EU through the collaborative
H2020 project QuProCS (Grant No. 641277) and by SERB through 
project VJR/2017/000011. MGAP is member of GNFM-INdAM.}

\end{document}